\documentclass[preprint, showpacs,preprintnumbers,aps,prd,amsmath,amssymb,nofootinbib]{revtex4}

\usepackage{graphicx}
\usepackage{dcolumn} 
\usepackage{bm}      

\newcommand{\bea}{\begin{eqnarray}}
\newcommand{\eea}{\end{eqnarray}}

\begin{document}
\title{Avoidance of Big Rip In Phantom Cosmology \\by Gravitational Back Reaction }         
\author{  Puxun Wu $^b$ and Hongwei Yu $^{a,b}$
\footnote{To whom correspondence should be addressed} }

\affiliation {  $ ^a$ CCAST(World Lab.), P. O. Box 8730, Beijing,
100080, P. R. China .
\\ $^b $Department of Physics and Institute of  Physics,\\ Hunan
Normal University, Changsha, Hunan 410081, China\footnote{Mailing
address .}}


\begin{abstract}
The effects of the gravitational back reaction of cosmological
perturbations are investigated in a cosmological model where the
universe is dominated by phantom energy. We assume a COBE
normalized spectrum of cosmological fluctuations at the present
time and calculate the effective energy-momentum tensor of the
gravitational back-reactions of cosmological perturbations whose
wavelengths at the time when the back-reactions are evaluated are
larger than the Hubble radius. Our results reveal that the effects
of gravitational back-reactions will counteract that of phantom
energy sooner or later and can become large enough to terminate
the phantom dominated phase before the big rip as the universe
evolves. This arises because the phase space of infrared modes
grows very rapidly as we come close to the big rip.
\end{abstract}


 \maketitle

\section{Introduction}
The analysis of data from supernovae\cite{Per}, CMB\cite{Spe} and
WMAP\cite{Bah} strongly indicates the existence of dark energy
which dominates the present universe and drives the accelerating
cosmic expansion. Many  models, such as cosmological
constant\cite{Wein}, quintessence\cite{Peeb},
k-essence\cite{kessen}, braneworld\cite{Brane}, Chaplygin
gas\cite{Chaply}, quintom\cite{quintom}, and
holography\cite{holo}, are proposed to explain the dark energy. A
phantom field \cite{Cal}, which has the super-negative equation of
state($ w \equiv p/\rho<-1$) in contrast to quintessence energy
($w>-1$) or cosmology constant($w=-1$), appears as another
possible candidate of dark energy and has received increased
attention recently. It has an unusual kinetic term in its
Lagrangian which gives rise to some strange properties, such as
the violation of the dominant energy condition\cite{Carr,Park} and
the increase of its energy density with time. As bizarre as it may
appear, such terms could arise in a variety of theories
\cite{Nill}.

It can be shown that once our  universe enters the phantom energy
dominated phase, the scalar factor will blow up in a finite proper
time due to excessive expansion. This arises because \cite{Cal,
Cald} the energy density of phantom fields increases with time
instead of red-shifting away as the matter or radiation energy
densities or as the energy density of ordinary quintessence. Thus
the increasing  phantom energy will ultimately strip apart
gravitationally bound bodies and cause a cosmic doomsday or big
rip\cite{Cald}.  The above conclusions are based upon a constant
negative value of $w$. However,  if the value of $w$ could change
during the evolution of the universe and then in principle the big
rip can be avoided \cite{Carr}. Attempts have also been made
toward avoiding the big rip by modifying the original Caldwell's
phantom model \cite{McIn,VS,PFGD,Bou,NDE}.  Note that it has been
argued that  a singularity can also develop at a finite future
time even if $\rho + 3p$ is positive \cite{Barrow}.

Here we would like to point out that the big rip may be avoided in
phantom cosmology even with a constant negative value of $w$
without adding new physics. We will demonstrate that the
gravitational back reaction of cosmological perturbations may
terminate the phantom dominance before the big rip occurs.

\section{Gravitational back reaction}

Considering the fluctuations of metric and matter  and expanding
the Einstein equation to  second order, due to non-linearity of
Einstein equation, the second order equation can not be satisfied
and the back reaction of these fluctuations to the background must
be  considered  which is characterized by a gauge-invariant
effective energy-momentum tensor $\tau_{\mu\nu}$\cite{Raul}
\begin{eqnarray}\label{1}
\tau_{\mu\nu}=\langle T_{\mu\nu}^{(2)}-\frac{1}{8\pi
G}{G_{\mu\nu}^{(2)}}\rangle,
\end{eqnarray}
where $T_{\mu\nu}^{(2)}$ and $G_{\mu\nu}^{(2)}$ express the second
order metric and matter perturbations and pointed brackets stand
for spatial averaging. This formalism can be applied  to both
scalar and tensor perturbations and applies independent of the
wavelength of the perturbations. The effects of gravitational back
reactions  have been studied in the context of cosmological
models, where, for example, they have been used to address the
issue of dynamical relaxation of  the cosmological constant
\cite{Bran} and the possible termination of quintessence
phase\cite{Li}.  Here we will discuss the back reactions in
phantom cosmology.

We will assume that $w$ is constant and $w<-1$,  and the universe
is not phantom dominated  until the time $t_m$.  That is, $t_m$ is
the time of equal phantom energy density and matter energy
densities.  The universe is matter dominated if $t< t_m$ and is
phantom dominated if otherwise. In the following we discuss the
back reaction of cosmological perturbations during the phantom
dominated phase ($t\geq t_m$). However, if  phantom exists  all
the time,  the back reaction should too before $t_m$. Here we
assume that the back reaction is negligible or its effects are
reinforcing before the phantom dominated era , since if otherwise
the phantom phase could not occur.  As the amplitude of each
fluctuation mode is small, we need a very large phase space of
modes in order to produce any interesting effects. During the
phantom dominated phase,  both the scalar factor $a$ and expansion
rate $H$ increase with time, so Hubble distance decreases and the
phase space of infrared modes grows. Hence  we expect that the
effects of the back reaction of infrared modes will grow  and we
therefore only focus on the back reaction of infrared modes on the
evolution of the universe dominated by phantom energy.

A simple effective action of a phantom field can be expressed as
\begin{eqnarray}\label{2}
{\cal L}_{phan}=-\frac{1}{2}(\partial_\mu \varphi)^2-V(\varphi),
\end{eqnarray}
where $V(\varphi)$ is the potential of the phantom field. Thus we
can obtain the energy momentum tensor
\begin{eqnarray}\label{EMT}
T_{\mu\nu}=-\partial_\mu \varphi \partial_\nu \varphi
+g_{\mu\nu}\left[\;
\frac{1}{2}\partial^\alpha\varphi\partial_\alpha\varphi+V(\varphi)\right]\;.
\end{eqnarray}
In longitudinal gauge the  metric  with scalar perturbations can
be written as
\begin{eqnarray}\label{3}
ds^2=a^2(\eta)[(1+2\Phi)d\eta^2-(1-2\Phi)\delta_{ij}dx^i dx^j]\;,
\end{eqnarray}
where $\Phi$ is the Bardeen potential. In addition to the
geometrical perturbations, one must also consider  the
perturbations of  the phantom field $\delta\varphi$ during the
phantom dominated era. Expanding the Einstein equation to the
first order in $\Phi$ and $\delta\varphi$,  we obtain the
gauge-invariant equations of  motion for small perturbations
\begin{eqnarray}\label{4}
\Phi'+h\Phi=-4\pi G \varphi_0'\delta\varphi\;,
\end{eqnarray}
\begin{eqnarray}\label{5}
\nabla^2\Phi-3h(h\Phi+\Phi')=4\pi G
[\varphi_0'^2\Phi-\varphi_0'\delta\varphi'+V,_\varphi
a^2\delta\varphi]\;,
\end{eqnarray}
\begin{eqnarray}\label{6}
\Phi''+3h\Phi'+(2h'+h^2)\Phi
  =-4\pi G
[-\varphi_0'^2\Phi+\varphi_0'\delta\varphi'+V,_\varphi
a^2\delta\varphi]\;,
\end{eqnarray}
where $a'$ represents the derivative with respect to conformal
time $\eta$ and $h=a'/a$.
Eqs.(\ref{4},\ref{5},\ref{6}) come from the $0i$, $00$ and $ij$
components  of the Einstein equation respectively.
 Subtracting
Eq.~(\ref{5}) from Eq.~(\ref{6}) and using Eq.~(\ref{4})  and the
equation of motion for the phantom field, we get a second order
partial differential equation for $\Phi$
\begin{eqnarray}\label{8}
\Phi''-\nabla^2\Phi+2\left(h-\frac{\varphi_0''}{\varphi_0'}\right)\Phi'+2\left(h'-
h\frac{\varphi_0''}{\varphi_0'}\right)\Phi=0\;.
\end{eqnarray}
This equation is the same as that in which the matter
perturbations are induced by a usual scalar field \cite{Mukh}.
Thus, for long-wavelength perturbations we have
\begin{eqnarray}\label{9}
\Phi_k\simeq A_k\left[1-\frac{H}{a}\int a dt\right]\;.
\end{eqnarray}
Here $A_k$ is an integration constant. For short-wavelength
perturbations one gets
\begin{eqnarray}\label{10}
\Phi_k\propto\dot{ \varphi}_0\;,
\end{eqnarray}
where a dot denotes the derivative with respect to coordinate time
$t$.

Expansion of the energy-momentum tensor $T_{\mu\nu}$ and Einstein
tensor $G_{\mu\nu}$  to the second order in $\Phi$ and
$\delta\varphi$ yields the non vanishing components of the
effective back reaction energy-momentum tensor
\begin{eqnarray}\label{11}
\tau_{00}&=&\frac{1}{8\pi
G}[12H\langle\Phi\dot{\Phi}\rangle-3\langle\dot{\Phi}^2\rangle+9a^{-2}\langle\nabla
\Phi^2\rangle] -\frac{1}{2}\langle\delta\dot{\varphi}^2\rangle
\nonumber\\
&&\;-\frac{1}{2}a^{-2}\langle(\nabla\delta\varphi)^2\rangle
+\frac{1}{2}V,_{\varphi\varphi}\langle\delta\varphi^2\rangle+2V,_\varphi\langle\Phi\delta\varphi\rangle\;,
\end{eqnarray}
and
\begin{eqnarray}\label{12}
\tau_{ij}&=&a^2 \delta_{ij}{\biggl\{}\frac{1}{8\pi G}{\biggl
[}(24H^2+16
\dot{H})\langle\Phi^2\rangle+24H\langle\Phi\dot{\Phi}\rangle
 +\langle\dot{\Phi}^2\rangle+4\langle\Phi\ddot{\Phi}\rangle-
\frac{3}{4}a^{-2}\langle\nabla \Phi^2\rangle{\biggl ]}\nonumber\\
&&
-4\dot{\varphi}_0^2\langle\Phi^2\rangle-\frac{1}{2}\langle\delta\dot{\varphi}^2\rangle
-\frac{1}{2}a^{-2}\langle(\nabla\delta\varphi)^2\rangle+4\dot{\varphi}_0\langle\Phi\delta\dot{\varphi}\rangle
\nonumber\\
&&-\frac{1}{2}V,_{\varphi\varphi}\langle\delta\varphi^2\rangle+2V,_\varphi\langle\Phi\delta\varphi\rangle{\biggl\}}\;,
\end{eqnarray}
where  $H=\dot{a}/a$.


Let us now apply these equations to the  phantom dominated era.
For $t>t_m$ the solution for the scale factor is given by

\bea\label{13}
a(t)=a(t_m)\bigg[-w+(1+w)\frac{t}{t_m}\bigg]^\frac{2}{3(1+w)}\;.
\eea Apparently $a$ diverges when $t=w t_m/(1+w)\equiv t_{brip}$.
During the phantom dominated era the energy density evolves as
$\rho_{phan}\sim a^{-3(1+w)}$, thus $\rho_{phan}(t)$ is related to
$\rho(t_m)$ by

\bea\label{14}
\rho_{phan}(t)=\frac{\rho(t_m)}{[-w+(1+w)t/t_m]^2}\;. \eea
  Here $\rho(t_m)\approx 1/(6\pi G t_m^2)$.
 The energy density $\rho_{phan}(t)$ increases with time and
diverges when $t= t_{brip}$. Using $p=w\rho$,
$p=-\frac{1}{2}\dot{\varphi}^2-V(\varphi)$ and
$\rho=-\frac{1}{2}\dot{\varphi}^2+V(\varphi)$,  we obtain

\bea\label{15}
\dot{\varphi}^2=\frac{-(w+1)\rho(t_m)}{[-w+(1+w)t/t_m]^2}\;, \eea
 and
\bea\label{16}
V(\varphi)=\frac{(-w+1)\rho(t_m)}{2[-w+(1+w)t/t_m]^2}=
\frac{(-w+1)\rho(t_m)}{2}\exp\left[\frac{-2(1+w)\varphi}{t_m\sqrt{-(w+1)\rho(t_m)}}\right].
\eea
 Combining Eq.~(\ref{9}) and Eq.~(\ref{13}),  we get in the
long-wavelength limit
\begin{eqnarray}\label{17}
\Phi_k=\bigg\{\begin{array}{ll}\quad\qquad\beta A_k=\tilde{A}_k,
\quad\qquad\;\;\;{w\neq-5/3}\;,\\
A_k[1-\ln \chi(t)]\equiv A_k z(t),\;\;{w=-5/3}\;,\end{array}
\end{eqnarray}
where $\beta$ is an integration constant and
$\chi(t)\equiv\frac{5}{3}-\frac{2t}{3t_m}$. As the universe
evolves from $t_m$ to $t_{brip}$ ,   $\chi(t)$ varies from $1$ to
$0$ and $z(t)$ from $1$ to $\infty$. It then follows that
$\dot{\Phi}_k=\ddot{\Phi}_k=0$ for $w\neq-5/3$ and
   \bea
\label{dp}\dot{\Phi}_k=
 \frac{2}{3t_m\chi(t)z(t)}\Phi_k\;,\eea
 and
\bea\label{ddp} \ddot{\Phi}_k=
 \frac{4}{9t_m^2\chi(t)^2z(t)}\Phi_k\;,\eea
 for $w=-5/3$ .
 Defining $\rho_{br}\equiv\tau^0_0$ and $p_{br}\equiv-\frac{1}{3}\tau^i_i$ and using Eqs.~(\ref{4},\ref{15},\ref{16},\ref{dp},\ref{ddp}),
the expression for $\tau_{\mu\nu}$ can be simplified to
\begin{equation}\label{18}
\rho_{br}=\bigg\{\begin{array}{ll}
\quad -\frac{(1-w)}{6\pi
Gt_m^2[-w+(1+w)t/t_m]^2}\langle\Phi^2\rangle,\quad\quad{w\neq-5/3}\;,\\
\frac{2}{9\pi Gt_m^2\chi(t)^2}
   \bigg(\frac{1}{z(t)^2}+\frac{3}{z(t)}
   -2\bigg)\langle\Phi^2\rangle,\;{w=-5/3}\;,\end{array}%
\end{equation}
\begin{equation}\label{19}
p_{br}=\bigg\{\begin{array}{ll}\qquad\qquad-\rho_{br},
\;\;\quad\qquad\qquad\qquad\qquad{w\neq-5/3}\;,\\
 -\frac{2}{9\pi Gt_m^2\chi(t)^2}
   \bigg(\frac{2}{z(t)^2}+\frac{3}{z(t)}
   -2\bigg)\langle\Phi^2\rangle,\;\quad{w=-5/3}\;.\end{array}
\end{equation}
Here the subscript $br$ stands for back reaction.
So the equation of state parameter for
the dominant infrared contribution to the back reaction is given
by
 \bea
 w_{br}=\bigg\{\begin{array}{ll}\qquad  -1,
 \quad\qquad\quad{w\neq-5/3}\;,
 \\-\frac{2+3z(t)-2z(t)^2}{1+3z(t)-2z(t)^2},\;\quad{w=-5/3}\;.\end{array}
   \eea
Hence, for $w\neq-5/3$, the contribution of infrared modes to the
energy momentum tensor which
   describes the back-reaction
   takes the form of a negative cosmological constant, whose absolute value changes
   as a function of time. The result of this case is similar to that obtained in an
   inflationary background cosmology \cite{Abramo,Brandenberger}, which has been used to address the
issue of dynamical relaxation of the cosmological constant
\cite{Bran}.
   Meanwhile,  for $w=-5/3$,  $w_{br}=-3/2$ at
   $t=t_m$,  and $w_{br}$ grows as time goes on and approaches to $-1$ at $t_{brip}$.
     Using Eq.~(\ref{14},
\ref{18}), we have
\begin{eqnarray}\label{bb}
\frac{\rho_{br}}{\rho_{phan}} =\bigg\{
\begin{array}{ll}\qquad
 -(1-w)\langle\Phi^2\rangle,\;\;\quad\qquad{w\neq-5/3}\;,\\
 \frac{4}{3}\bigg(\frac{1}{z(t)^2}+\frac{3}{z(t)}
   -2\bigg)\langle\Phi^2\rangle,\;\quad{w=-5/3}\;.\end{array}
\end{eqnarray}
  If the above ratio is negative, the phantom
energy will be counteracted by the effects of the back reaction.
When the ratio becomes negative unit, the phantom phase  will end
and the universe will re-enter the matter dominated era.  In order
to determine the value of this ratio, it is pivotal to evaluate
the two-point function $\langle \Phi^2\rangle$, which can be
obtained by integrating over all Fourier modes of $\Phi$:
\begin{eqnarray}\label{23}
\langle\Phi(t)^2\rangle=\int_{k_i}^{k_t}
\frac{k^3}{2\pi^2}\langle\Phi_k(t)^2\rangle\frac{dk}{k}\;,
\end{eqnarray}
where $k_i=a_i H_i$ and $k_t=a(t)H(t)$ are infrared and
ultraviolet cutoffs respectively. The infrared cutoff can be
chosen as the length scale above which there are no significant
fluctuations. If we assume that there was a period of inflation in
the early history of our universe, we could take the infrared
cutoff as the Hubble radius at the beginning of the inflation. The
ultraviolet cutoff will be taken as the Hubble radius at time $t$
when the strength of the back reaction is to be evaluated.

\section{Termination of phantom dominated phase}

We assume that our universe now is dominated by the phantom
energy, and for simplicity, $t_m\sim t_0$, which means that the
phantom just begins to dominate at today. As a matter of fact, one
of the main features that distinguish the phantom energy from a
cosmological constant or quintessence is that the onset of phantom
energy dominance happens at the very last moment. So our
assumption is fairly reasonable. Now we want to relate $\Phi_{pk}$
defined as the value of $\Phi_{k}$ during the phantom dominated
phase, and $\Phi_{mk}$, which is the corresponding value just
before the phantom energy begins to dominate. Let us consider
separately modes whose wavelengths are outside the Hubble radius
now and modes whose wavelengths are inside the Hubble radius today
and exit the Hubble radius before time $t$($t>t_0$). For the
former case,  $\Phi_{pk}(t)$ and $\Phi_{mk}(t_0)$ can be related
by the conservation of the Bardeen potential\cite{Mukh, Bard}
\begin{eqnarray}
\zeta=\frac{2}{3}\frac{(H^{-1}\dot{\Phi}+\Phi)}{1+w}+\Phi\;.
\end{eqnarray}
In the matter dominated phase, since $w=0$ and
$\dot{\Phi}_{mk}=0$, we obtain $\zeta (\Phi_{mk})=5\Phi_{mk}/3$.
For $w\neq-5/3$,  during the phantom dominated era,
$\dot{\Phi}_{pk}=0$. So we have  $\zeta
(\Phi_{pk})=(5+3w)/(3+3w)\Phi_{pk}$.  It follows from the
conservation of  $\zeta$  that
\begin{eqnarray}\label{pmk2}
\Phi_{pk}=\frac{5(1+w)}{3w+5}\Phi_{mk}\;.
\end{eqnarray}
For modes whose  their wavelengths  are inside the Hubble radius
today and exit the Hubble radius at time $t_H$ given by
$a(t_H)H=k$. From Eqs.~(\ref{10},\ref{15},\ref{17}), we know how
 $\Phi$ evolves with time on the wavelength scale smaller and larger than
the Hubble radius. Thus we can express $\Phi_{pk}(t>t_H)$ in terms
of $t_0$, $t_H$ and $\Phi_{mk}(t_0)$
\begin{eqnarray}\label{25}
\Phi_{pk}(t)=\Phi_{mk}(t_0)\frac{t_0}{[-w t_0+(1+w)t_H]}\;.
\end{eqnarray}
For the case of $w=-5/3$,  we
 have in the matter dominated era, from Eq.~(\ref{9}), that $\Phi_{mk}=\frac{3}{5}A_k$ in the long-wavelength
 limit. For modes whose wavelengths are
outside the Hubble radius now,  using  Eq.~(\ref{17}), we
 have
 \bea\label{Phik1}
    \Phi_{pk}(t)=\frac{5}{3}\Phi_{mk}(t_0)z(t)\;.
 \eea
For those modes whose wavelengths are inside the Hubble radius
today and exit the Hubble radius before time $t$($t>t_0$), using
Eqs.(\ref{10},\ref{15},\ref{17}), we obtain
  \bea\label{Phik2}
  \Phi_{pk}(t)=\Phi_{mk}(t_0)\frac{\chi(t_0)}{\chi(t_H)}\frac{z(t)}{z(t_H)}\;.\eea

We now assume that the spectrum of cosmological fluctuations at
the present time is normalized by the recent observations of CMB
anisotropies and  the fluctuational spectrum has the form
\begin{eqnarray}\label{26}
P(k)\equiv
\frac{k^3}{2\pi^2}|\Phi_{mk}(t_0)|^2=C\left(\frac{k}{k_{COBE}}\right)^\lambda\;,
\end{eqnarray}
where $k_{COBE}\equiv \alpha a_0 H_0$ and $C^{1/2}\simeq 10^{-5}$.
By the joint analysis of the Maxima-1, Boomerang and COBE cosmic
microwave anisotropy results and the estimate of the systematic
errors, the blue spectrum tilt should be in the region of
$0\leq\lambda<0.27$ \cite{Jaff}\footnote{According to the more
recent WMAP data the upper bound could be slightly larger
\cite{Peiris,Seljak}.}. In this case $\alpha\simeq 7.5$.

\begin{figure}
\includegraphics{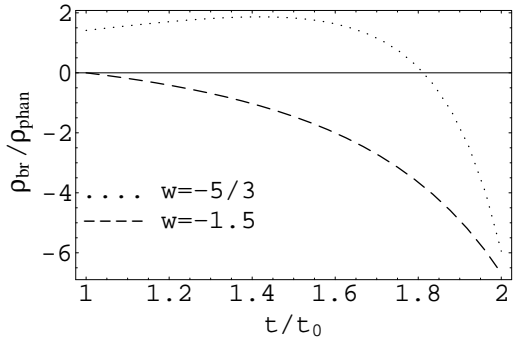}
\caption{\label{fig:1} $\rho_{br}/\rho_{phan}$ vs $t/t_0$ is
plotted for $w=1.5$ and $w=-5/3$ at the early epoch of phantom
dominated phase with $\lambda=0.27,\alpha=7.5, C=10^{-10}$. The
vertical coordinate has been scaled by $C=10^{-10}$.}
\end{figure}

\begin{figure}
\includegraphics{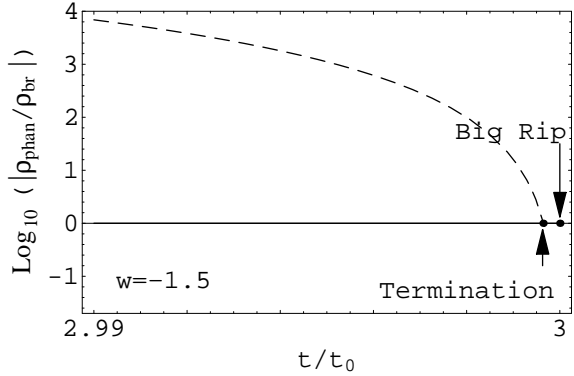}
\caption{\label{fig:2} Plotted is
$Log_{10}(|\rho_{phan}/\rho_{br}|)$ vs $t/t_0$ for  $w=-1.5$ with
$ \lambda=0.27,\alpha=7.5, C=10^{-10}$. The step length of the
horizontal axis is $0.0005$. Note that $t_0\approx $15Gyr }
\end{figure}

\begin{figure}
\includegraphics{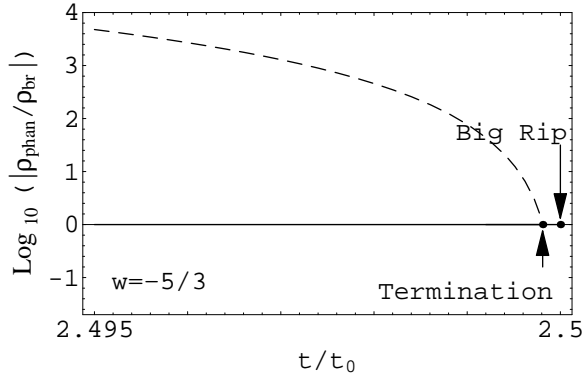}
\caption{\label{fig:3} $Log_{10}(|\rho_{phan}/\rho_{br}|)$ vs
$t/t_0$ is plotted for $w=-5/3$ with $\lambda=0.27,\alpha=7.5,
C=10^{-10}$. The step length of the horizontal axis is $0.0002$.
Note that $t_0\approx $15Gyr}
\end{figure}

 Substituting Eqs.~(\ref{pmk2},\ref{25},\ref{26}) into
Eq.~(\ref{23}), we obtain, $w\neq-5/3$
\begin{widetext}
\begin{eqnarray}\label{29}
\langle\Phi(t)^2\rangle  \simeq \frac{25C(w+1)^2}{(3w+5)^2(\alpha
a_0 H_0)^{\lambda}}\int_{k_i}^{k_0}k^{\lambda-1}dk +
\frac{Ct_0^2}{(\alpha a_0H_0)^{\lambda}
}\int_{k_0}^{k_t}\frac{k^{\lambda-1}}{[-w t_0+(1+w)t_H]^2}dk\;.
\end{eqnarray}
 The first integration in the above  can be evaluated  as
\bea
 g_1&\equiv & \frac{25C(w+1)^2}{(3w+5)^2(\alpha a_0
H_0)^{\lambda}}\int_{k_i}^{k_0}k^{\lambda-1}dk\nonumber\\
&=&\frac{25C(w+1)^2}{(3w+5)^2}f_1\;,
  \eea
  where we have defined
  \begin{equation}
f_1=\bigg\{\begin{array}{ll}
\qquad\ln\frac{a_0H_0}{k_i},\;\qquad\qquad\quad\;{
\lambda=0}\;,\\
\lambda^{-1}\alpha^{-\lambda}[1-k_i^\lambda  (a_0
H_0)^{-\lambda}],\; {\lambda\neq 0}\;.\end{array}
  \end{equation}
  For an e-folding number as large as $60\sim 70$,
$\ln(a_0H_0/k_i)\sim {\cal O}(10^1)$. Therefore, since $C\sim
10^{-10}$,   $g_1\ll 1$. The second integration is given by
\bea\label{g2} g_2&\equiv&\frac{Ct_0^2}{(\alpha a_0H_0)^{\lambda}
}\int_{k_0}^{k_t}\frac{k^{\lambda-1}}{[-w t_0+(1+w)t_H]^2}dk
\nonumber\\
&=&\frac{(1+3w)}{6(1+w)+(1+3w)\lambda}\frac{C}{\alpha^\lambda}
\bigg(\left[-w+(1+w)\frac{t}{t_0}\right]^{-\frac{1+3w}{3(1+w)}\lambda-2}
 -1\bigg)\;.
   \eea
Apparently $g_2\sim 0$ when $t\sim t_0$ and  it increases with
time and approaches infinity at $t=t_{brip}$.   Thus hereafter, we
will discard $g_1$ in discussing if and when the effects of back
reaction can terminate the phantom dominated phase before
everything is torn apart at the big rip.

For $w=-5/3$,  the substitution of
Eqs.~(\ref{Phik1},\ref{Phik2},\ref{26}) into Eq. (\ref{23}) leads
to
 \bea\label{Phit2}
  \langle\Phi(t)^2\rangle&=&\frac{25Cz(t)^2}{9k_{COBE}^\lambda}\int_{k_i}^{k_0}k^{\lambda-1}dk+
  \frac{Cz(t)^2}{k_{COBE}^\lambda}\int_{k_0}^{k_t}\frac{k^{\lambda-1}}
  {z(t_H)^2\chi(t_H)^2}dk\nonumber\\
  &=&2Cz(t)^2\left[\frac{25}{18}f_1+f_2(t) \right]\;,\eea
  where
\bea
 f_2(t)&\equiv&\frac{1}{2k_{COBE}^\lambda}\int_{k_0}^{k_t}\frac{k^{\lambda-1}}
  {z(t_H)^2\chi(t_H)^2}dk\nonumber\\
  &=& 2\alpha^{-\lambda}(1+\lambda)e^{-2(1+\lambda)}Ei[2(1+\lambda)z(t)]
  -\frac{\chi(t)^{-2(1+\lambda)}}{\alpha^\lambda z(t)}\;.\eea
Here $Ei(x)$ is the exponential integral function. Appealing to
the fact that when $x\rightarrow \infty$
 $Ei(x)$ can be expanded as
  \bea
Ei(x)\sim e^x(x^{-1}+x^{-2}+\cdots)\;,
 \eea
we have that
 \bea
f_2(t)\simeq
  \alpha^{-\lambda}\chi(t)^{-2(1+\lambda)}\left[\frac{z(t)^{-2}}{2(1+\lambda)}+\cdots\right],
  \eea
  when $t$ approaches $t_{brip}$.
The above result blows up at the big rip, so  we will discard
$\frac{25}{18}f_1$ in Eq.~(\ref{Phit2}).  Using
Eqs.~(\ref{bb},\ref{29},\ref{g2},\ref{Phit2}), we obtain
\begin{eqnarray}\label{brp}
\frac{\rho_{br}}{\rho_{phan}} \simeq\bigg\{\begin{array}{ll}
-\frac{(1+3w)(1-w)}{6(1+w)+(1+3w)\lambda}\frac{C}{\alpha^\lambda}
\bigg(\left[-w+(1+w)\frac{t}{t_0}\right]^{-\frac{1+3w}{3(1+w)}\lambda-2}
 -1\bigg),\; {w\neq-5/3}\;,\\
\qquad\qquad\qquad \frac{8C}{3}[1+3z(t)-2z(t)^2]
f_2(t),\qquad\qquad\qquad\;{w=-5/3}\;.\end{array}
\end{eqnarray}
It is interesting to note that when $w\neq -5/3 $, the ratio is
always negative as long as $w<-1$ and becomes
 negative unity before $t\rightarrow t_{brip}$.
 While, for the case of $w=-5/3$, it can be shown that the ratio is positive at
 present and remains so for a period of time.  The plot of this ratio vs $t/t_0$ in Fig.~{\ref{fig:1}} shows that it
 turns negative approximately at $t\approx 1.81t_0$.  This indicates that the  back reaction reinforces the
phantom energy in an early period of phantom dominated universe
and then counteracts it. At the big rip this ratio becomes
negative infinity.  Therefore, the phantom dominated phase will be
terminated sooner or later before the big rip by the back reaction
effects.  The behaviors of the energy density of the back reaction
and that of the phantom background as a function of $t/t_0$ are
plotted in Fig.~\ref{fig:2} and Fig.~\ref{fig:3}. There we can see
that the ratio of $|\rho_{phan}/\rho_{br}|$ reaches unity (i.e.,
$Log_{10}(|\rho_{phan}/\rho_{br}|)$ becomes zero) before the big
rip and thus the phantom phase terminates. Note that the
termination point appears very close to the big rip in the
figures. However, since $t_0$($\approx$15Gyr) a is very large
number, the actual time interval measured in terms of years is not
small (see also the Table).
 Let us now discuss in more detail when
 this happens.
For the case of $w\neq-5/3$ the time when the phantom phase is
terminated by the effects of gravitational back reaction, i.e.,
when $\frac{\rho_{br}}{\rho_{phan}} \simeq -1$,  can be explicitly
given as
\begin{eqnarray}\label{end}
t\simeq
\frac{t_0}{1+w}\bigg(\left[\frac{6(1+w)+(1+3w)\lambda}{(1+3w)(1-w)}
\frac{\alpha^\lambda}{C}+1\right]^{-\frac{3(1+w)}{(1+3w)\lambda+6(1+w)}}+w\bigg)
\;.
\end{eqnarray}
Define  $t'$ as the time of termination of the phantom phase
before the big rip,  we have
\begin{eqnarray}
\label{Termi}
 t'\equiv
t_{brip}-t=-\frac{t_0}{1+w}\left[\frac{6(1+w)+(1+3w)\lambda}{(1+3w)(1-w)}
\frac{\alpha^\lambda}{C}+1\right]^{-\frac{3(1+w)}{(1+3w)\lambda+6(1+w)}}
\;.
\end{eqnarray}\end{widetext}
For a fixed $w$, $t'$  is an increasing function of $\lambda$.
This is similar to
 the case of quintessence\cite{Li} in that the effect of back reaction is
 proportion to the blue tilt.  For scale invariant spectra, i.e.,
 $\lambda=0$ ,
 the effect of back reaction for quintessence is
 negligible\cite{Li}. In contrast, for the phantom case, it is
 easy to see  from Eq.~(\ref{Termi}) that
 the effect of back reaction can still become large enough to terminate the phantom
 phase.
When $w=-5/3$,  it is almost impossible to solve Eq.(\ref{brp}) to
get an analytical expression for the  time when
$\rho_{br}/\rho_{phan}\simeq-1$,  but we can  resort to numerical
techniques.    Listed in the Table are  results of the
calculations for the time of termination of the phantom dominated
phase for both $w\neq -5/3$ and $w=-5/3$.

 \begin{table}
\begin{ruledtabular}\caption{\label{tab:table1} The time  of termination  of the
phantom phase before the big rip:
$t_0=15Gyr,\alpha=7.5,C=10^{-10}$.}
\begin{tabular}{lcr}
w& $t'(yr)(\lambda=0)$ & $t'(yr)(\lambda=0.27)$\\
\hline
-1.1 & $4.2\times 10^6$ & $6.4\times 10^8$\\
-1.3 & $9.6\times 10^5 $ & $1.9\times 10^7$\\
-1.5 & $5.1\times 10^5 $& $5.2\times 10^6$\\
-5/3 & $3.4\times 10^5$ & $2.7\times 10^6 $\\
-1.8 & $3 \times 10^5$ & $2\times 10^6$\\
-2   & $2.4 \times 10^5$ & $1.3 \times 10^6$\\
\end{tabular}
\end{ruledtabular}
\end{table}

\section{Discussions}

 The main conclusion  we can draw  from  the Table is that
the big rip  can be avoided by the gravitational back reaction of
cosmological perturbations. A comparison of our results with that
of Ref.~\cite{Cald} for the case of $w=-1.5$ shows that our Solar
system could be saved!. We can also see from the table that for a
fixed $w$ the greater the blue tilt, $\lambda$, the sooner the
termination of the phantom phase.   However, for a fixed
$\lambda$, the greater  the  $w$,  the earlier the termination of
phantom phase before the big rip.  The physical reason is that the
greater  the $w$, the farther away we are from the big rip  and
thus there is more time for  the infrared modes  to accumulate.
Finally, let us consider a case of pure theoretical significance,
i.e., the case in which $\lambda\rightarrow \infty$, then we find
$t'\simeq 13Gyr$ for $w\neq-5/3$ and $t_0=15Gyr$ .  This means
that it is not possible to kill the phantom dominated phase the
time just when it kicks in.

In summary,  We have, assuming a COBE normalized spectrum of
cosmological fluctuations at the present time, calculated the
gravitational back-reaction effects of cosmological perturbations
whose wavelengths at the time when the back-reactions are
evaluated are larger than the Hubble radius. Our results reveal
that the gravitational back-reactions are growing with time and
could become large enough to terminate the phantom dominated phase
before the big rip occurs. An interesting feature to be noted of
the gravitational back reactions is that their effective energy
momentum tensor is that of some form of "matter" which has
negative energy density and positive pressure.  This form of
"matter" was postulated in Ref.~\cite{McIn}

\begin{acknowledgments}
This work was supported in part  by the National Natural Science
Foundation of China  under Grant No. 10375023, the Program for
NCET (No. 04-0784), the Key Project of Chinese Ministry of
Education (No. 205110) and the National Basic Research Program of
China under Grant No. 2003CB71630.
\end{acknowledgments}

\end{document}